\documentclass[copyright,noderivs]{eptcs}
\providecommand{\event}{AFL 2026} % Name of the event you are submitting to
\usepackage{underscore}           % Only needed if you use pdflatex.

\usepackage{graphicx}
\usepackage{mathrsfs}
\usepackage{amssymb}
\usepackage{amsmath}
\usepackage{amsthm}
\usepackage{textcomp}
\usepackage{multirow}
\usepackage{comment}

\newtheorem{theorem}{Theorem}

\newtheorem{corollary}[theorem]{Corollary}

\newcommand{\compl}[1]{\overline{#1}}

\newcommand{\fin}{\textrm{FIN}}
\newcommand{\cofin}{\textrm{COFIN}}
\newcommand{\ukf}{\textrm{UKF}}

\newcommand{\ufam}{\mathscr{U}}
\newcommand{\regex}{\textrm{\upshape{RE}}}

\newcommand{\reg}{\textsf{REG}}

\usepackage{tikz}
\usetikzlibrary{positioning,shadows.blur,fit,backgrounds}
\title{Subregular Expressions and Their Expressive Power}

\author{Martin Kutrib
\institute{%
  Institut f{\"u}r Informatik, Universit{\"a}t Giessen\\
  Arndtstr.~2, 35392 Giessen, Germany}
\email{kutrib@uni-giessen.de}
\and 
Matthias Wendlandt
\institute{%
  Institut f{\"u}r Informatik, Universit{\"a}t Giessen\\
  Arndtstr.~2, 35392 Giessen, Germany}
\email{wendlandt@uni-giessen.de}}

\def\titlerunning{Subregular Expressions and Their Expressive Power}
\def\authorrunning{M.~Kutrib, M.~Wendlandt}

\begin{document}

\maketitle

\begin{abstract}
We provide a survey of several families of subregular expressions obtained by modifying the
classical operator set consisting of union, concatenation, and Kleene
star. More specifically, we consider expression systems based on sets of two
or three operations chosen from union, intersection, complementation,
concatenation, and Kleene star. The resulting language families range from
finite languages to the full class of regular languages. We collect and
organize known results on the expressive power and structural properties of
these families. We compare the resulting language classes over unary and
arbitrary alphabets and summarize known inclusion, equality, and
incomparability results. 

The aim of this survey is to provide a unified view of the language families
induced by restricted operator sets and to highlight open relations and
possible directions for further research. 
\end{abstract}

\section{Introduction}
Regular languages admit a variety of equivalent formalisms, among which
regular expressions are one of the most widely used. Traditionally, they
are built from the basic operations of union, concatenation, and Kleene
star. Beyond these core operations, the class of regular languages is
closed under additional operations such as intersection and
complement, that can be applied without increasing the expressive power
\cite{Hopcroft:1979:itatlc:book}. This observation has motivated extensive
research on alternative expression systems and on the influence that the
choice of available operators has on expressive power and succinctness.

One prominent line of research considers extensions of classical regular
expressions. Modern regular expression engines frequently support
additional features such as variables and backreferences, yielding
\emph{extended regular expressions} (also known as practical regular
expressions or regex). In contrast to classical regular expressions,
these formalisms are capable of describing certain non-regular languages
\cite{Freydenberger:2013:eresd}. This increase in expressive power comes at a
considerable computational cost: fundamental decision problems such as
universality, equivalence, inclusion, regularity, and cofiniteness become
undecidable, even when only a single variable is available
\cite{Freydenberger:2013:eresd}. Nevertheless, important subclasses remain
tractable; for example, polynomial-time matching algorithms are known for
large classes of extended regular expressions
\cite{Reidenbach:2010:ptmt}. Related investigations of expressive power
and decidability for extended regular expressions can also be found in
\cite{Carle:2099:ere}.

Another natural direction is not to extend but to restrict the available
operations. A classical example is given by the star-free languages,
which are obtained by excluding the Kleene star while allowing union,
concatenation, and complement~\cite{cohen:1971:ddosfe}. Although still capable of
describing infinite languages, this class is strictly less expressive
than the full class of regular languages. It has been studied intensively
and admits elegant characterizations in algebraic and automata-theoretic
terms, for instance via aperiodic syntactic
monoids~\cite{Schuetzenberger:1965:ofmhots}, permutation-free deterministic
automata~\cite{McNaughton:1971:cfa:book}, and restricted forms of
alternating automata~\cite{Salomaa:2000:afasfl:art}.

Another natural restriction arises when the union operation is omitted.
The resulting class of \emph{union-free languages} has been studied
extensively and forms one of the best understood proper subclasses of
the regular languages. Besides its syntactic definition by regular
expressions without union, the class admits an elegant
characterization. It coincides with the family of
languages accepted by finite automata having exactly one cycle-free
accepting path from every state (so-called $1$CFPAs)
\cite{nagy:2004:nfre:proc, nagy:2006:ufrlcfpa}. Furthermore, variants such as deterministic
union-free languages and $\lambda$-free nondeterministic union-free
languages give rise to a proper hierarchy of increasingly expressive
language families \cite{Jiraskova:2012:ufdl, Nagy:2019:ufdc, Nagy:2021:ufr}.

Despite the absence of the union operation, union-free languages retain
a remarkably high expressive power. Every regular language can be
represented as a finite union of union-free languages, which motivates
the notion of \emph{union-complexity}, defined as the minimum number of
union-free components required in such a decomposition
\cite{nagy:2006:ufrlcfpa}. From an automata perspective, state complexity
questions for union-free languages have been investigated, showing that they
attain the same maximal complexity as general regular languages with respect
to several fundamental operations and complexity 
\mbox{measures~\cite{Brzozowski:2018:cdufrl,Jiraskova:2011:cufrl}.} 

In this survey, we place these results into a unified framework that
allows a systematic comparison with other restricted regular expression
formalisms.

Decision problems for restricted and extended variants of regular expressions
have also received considerable attention. Hunt~III showed strong lower bounds for
decision problems involving regular expressions with intersection, including
the equivalence problem \cite{hunt:1973:epreinpt}. More generally,
regular-like expressions with intersection are known to exhibit significantly
higher computational complexity than their classical counterparts. 

Investigations of these restricted and extended expression systems have
revealed a rich spectrum of complexity phenomena. In particular, the
computational difficulty of decision problems for regular and
regular-like expressions varies considerably depending on the available
\mbox{operations~
\cite{fuerer:1980:ciprei,hunt:1973:epreinpt,petersen:2000:dpfgre,petersen:2002:mprewic,stockmeyer:1974:cdpatl,Stockmeyer:1973:WPR}.}
An overview of many of these results is provided in
\cite{holzer:2011:crle:proc}. Together, these studies demonstrate that seemingly
minor modifications of the operator set can lead to substantial
differences in expressive power, structural properties, hierarchies,
succinctness, and computational complexity.

In this survey, we consider expression systems based on sets of two or three operations chosen from union, intersection, complementation, concatenation, and Kleene star. The resulting language families range from finite languages to the full class of regular languages. We collect and organize known results on their expressive power and structural properties, compare the resulting classes over unary and arbitrary alphabets, and summarize known inclusion, equality, and incomparability results.  Many results can be found in \cite{Kutrib:2025:seto,kutrib:2015:eccf:proc,kutrib:2017:cfl,kutrib:2018:ecse} in detail.

The remainder of the paper is organized as follows. After introducing the basic notation and the general framework for subregular expressions, we first investigate expression systems generated by sets of two operations. In particular, we consider the operation of union, intersection, complement, Kleene star and concatenation. We characterize the resulting language families, study their structural properties, and compare their expressive power over unary and arbitrary alphabets. We then turn to expression systems with three operations, where the interaction between the available operators leads to a more intricate inclusion structure and several open relations. Finally, we summarize the resulting landscape and outline several directions that may be of interest for future research, including questions concerning expressive power, decidability, structural characterizations, and descriptional complexity.

\section{Definitions and Preliminaries}\label{sec:prelim}

We write $\Sigma^*$ for the set of all words over the finite
alphabet $\Sigma$, we denote the \emph{empty word} as $\lambda$ and let $\Sigma^+= \Sigma^*\backslash\{\lambda\}$.
Given a word $w \in \Sigma^*$, we say that $|w|$ is its \emph{length}.
We use $\subseteq$ for \emph{inclusions} and~$\subset$ for
\emph{strict inclusions}.
The \emph{complement} of a language $L$ over alphabet $\Sigma$ is
is again a language over alphabet~$\Sigma$ which is denoted by $\compl{L}$.
The \emph{family of finite languages} is denoted by $\fin$,
the family of cofinite languages by $\cofin$, and the family of regular languages by $\reg$.

The \emph{regular expressions} over an alphabet~$\Sigma$ and the languages they
describe are defined inductively in the usual way:
$\emptyset$ and every word (of length one) $v\in \Sigma$ are regular
expressions, and when~$s$
and~$t$ are regular expressions, then $(s\cup t)$, \mbox{$(s\cdot t)$,} and
$(s)^*$ are also regular expressions.  The language $L(r)$ defined by a
regular expression~$r$ is defined as follows:
$L(\emptyset)=\emptyset$, $L(v)=\{v\}$, \mbox{$L(s\cup t)=L(s)\cup L(t)$,}
$L(s\cdot t)=L(s)\cdot L(t)$, and $L(s^*)=L(s)^*$.

Since the regular languages are closed under many more operations,
the approach to add operations like intersection ($\cap$) and
complementation ($\compl{\phantom{m}}$) does not increase their
expressive power. However, replacing operations by others may decrease
the expressive power. So, in general,
$\regex(\Sigma,\Lambda,\Phi)$, where
$\Lambda \subset \Sigma^*$ is a finite set of initial words, and
$\Phi$ is a set of (regularity preserving) operations, denotes all
regular(-like) expressions over $\Lambda$ using only operations
from $\Phi$. The language class induced by $\Phi$ is defined by
\[
\mathcal{L}_{\Phi}
=
\{\,L \mid
\exists \alpha \in \regex(\Sigma,\Lambda,\Phi)\colon
L = L(\alpha)\,\}.
\]

For a set of operations
$\Phi=\{\varphi_1,\ldots,\varphi_k\}$,
we abbreviate $\mathcal{L}_{\Phi}$ by
$\mathcal{L}_{(\varphi_1,\ldots,\varphi_k)}$.

Hence,
$\regex(\Sigma,\Sigma,\{\cup,\cdot,*\})$ refers to the set of all
ordinary regular expressions, and
$\mathcal{L}_{(\cup,\cdot,*)}$ denotes the class of regular languages.
Likewise,
$\regex(\Sigma,\Sigma,\{\cup,\cdot,\compl{\phantom{m}}\})$ refers to the
set of star-free expressions, and
$\mathcal{L}_{(\cup,\cdot,\compl{\phantom{m}})}$ denotes the class of
star-free languages.

Since in the presence of concatenation, every 
word in $\Lambda$ can be obtained by concatenating letters from~$\Sigma$,
the words in $\Lambda$ can be created for free. Moreover, in the presence
of union, every finite subset of words in $\Lambda$ can be created
for free. Here, as in previous works
\cite{kutrib:2015:eccf:proc,kutrib:2017:cfl,kutrib:2018:ecse}, 
we do not have necessarily concatenation or union and, thus, 
provide initially \emph{finite subsets} of words as literals in order to allow 
non-trivial languages to be expressed. Moreover,
we do this for uniformity and comparability for all types
in question.

For convenience, parentheses in expressions from
$\regex(\Sigma,\Lambda,\Phi)$ are sometimes omitted. In this case, the
unary operations complementation and Kleene star are assumed to have
higher precedence than all binary operations.
\bigskip

Before investigating the language classes in detail, we first
consider those classes whose expressive power can be determined
immediately. By Kleene's theorem, the class
$\mathcal{L}_{(\cup,\cdot,*)}$ coincides with the class of regular
languages. Since the regular languages are closed under intersection and
complementation, every extension of this operator set by $\cap$ and/or
$\compl{\phantom{m}}$ yields the same language family. Hence,
$$
\mathcal{L}_{(\cup,\cdot,*)}
=
\mathcal{L}_{(\cup,\cdot,*,\cap)}
=
\mathcal{L}_{(\cup,\cdot,*,\compl{\phantom{m}})}
=
\mathcal{L}_{(\cup,\cdot,*,\cap,\compl{\phantom{m}})}
=
\mathsf{REG}.
$$

\section{Subregular Expressions with two Operations}

In this section, we consider expression systems based on sets of two operations. Restricting the available operator set to only two operations provides a particularly fine-grained perspective on the individual contribution and expressive power of the different operations. It allows us to isolate their roles more clearly and to study how specific pairs of operations interact. In this way, the resulting language classes provide a fundamental basis for understanding the expressive landscape of more general expression systems with larger operator sets.

Several classes can immediately be characterized. Since neither Kleene
star nor complementation is available in the classes
$\mathcal{L}_{(\cup,\cdot)}$,
$\mathcal{L}_{(\cup,\cap)}$, and
$\mathcal{L}_{(\cdot,\cap)}$, every expression starts from finite
languages and applies only operations preserving finiteness.
Consequently, all three classes coincide with the family of finite
languages:
\[
\mathcal{L}_{(\cup,\cdot)}
=
\mathcal{L}_{(\cup,\cap)}
=
\mathcal{L}_{(\cdot,\cap)}
=
\fin.
\]

Another class whose expressive power follows immediately from closure
properties is
$\mathcal{L}_{(\cup,\compl{\phantom{m}})}$.
Since union preserves finiteness, complementation exchanges finite and
cofinite languages, and the union of finite and cofinite languages is
again either finite or cofinite, every language generated by these two
operations is finite or cofinite. Conversely, finite languages are
available from the finite basis, and cofinite languages are obtained as
complements of finite languages. By De Morgan's laws, union and
intersection are interdefinable in the presence of complementation.
Hence,
\[
\mathcal{L}_{(\cup,\compl{\phantom{m}})}
=
\mathcal{L}_{(\cap,\compl{\phantom{m}})}
= \fin\cup\cofin.
\]

Thus, only five genuinely different classes remain to be considered:
\[
\mathcal{L}_{(\cup,*)},\qquad
\mathcal{L}_{(\cdot,*)},\qquad
\mathcal{L}_{(\cdot,\compl{\phantom{m}})},\qquad
\mathcal{L}_{(\cap,*)},\qquad
\mathcal{L}_{(*,\compl{\phantom{m}})}.
\]
These classes possess considerably richer structural properties \cite{Kutrib:2025:seto}. We
discuss their expressive power in the following.

We first consider the class
$\mathcal{L}_{(\cup,*)}$.
Since concatenation is not available, arbitrary words cannot be
constructed from the alphabet symbols alone. Therefore, the finite
initial set
$\Lambda\subseteq\Sigma^*$
of words plays an essential role in this setting. Starting from this
basis, expressions in
$\mathcal{L}_{(\cup,*)}$
are formed using only union and Kleene star.

The crucial
observation is that the Kleene star always repeats an entire
subexpression. Hence, it cannot in general isolate a fixed prefix or
separator from an independently iterated suffix. Therefore, even simple
regular languages of the form $uv^*$ or $v^*u$ may be impossible to
describe without concatenation.

\begin{theorem}
The language
\[
L=b(aa)^*
\]
is not contained in
$\mathcal{L}_{(\cup,*)}$.
\end{theorem}

Consequently, the class
$\mathcal{L}_{(\cup,*)}$
forms a proper subclass of the regular languages.

A natural question is how the expressive power changes if a bounded
number of concatenation operations is added to this language family.
Each additional concatenation makes it possible to combine one further
fixed component with independently repeated parts. Hence, expressions
with more concatenations can describe increasingly complex regular
languages.

In particular, languages of the form
\[
u_0v_1^*u_2v_3^*\cdots u_{k-1}v_k^*
\]
require a corresponding number of concatenation operations, provided
that the fixed separator words cannot be generated inside one of the
iterated factors. This yields an infinite hierarchy with respect to the
number of allowed concatenations.

\begin{theorem}
For every integer $k\geq 1$, the family of languages described by
expressions of $\mathcal{L}_{(\cup,*)}$ using at most $k-1$
concatenation operations is strictly contained in the family obtained by
allowing $k$ concatenations.
\end{theorem}

Finally, the structure of expressions in
$\mathcal{L}_{(\cup,*)}$
is surprisingly simple with respect to the nesting depth of Kleene
stars. Whereas it is a classical question for regular expressions how
expressive power depends on star height, no such hierarchy arises here.
Since the only available operations are union and Kleene star, nested
stars can always be flattened.

\begin{theorem}
Every language in
$\mathcal{L}_{(\cup,*)}$
is either finite or can be represented as
\[
L_1^*\cup L_2^*\cup\cdots\cup L_k^*,
\]
where each $L_i$ is finite.
\end{theorem}

Among the remaining classes, the best understood one is
$\mathcal{L}_{(\cdot,*)}$, which is closely related to the family of
union-free languages. This class has been investigated extensively by
Nagy and coauthors, who established a rich collection of structural,
automata-theoretic, and complexity-theoretic results, including several
equivalent characterizations and hierarchy theorems 
\cite{Brzozowski:2018:cdufrl,Jiraskova:2011:cufrl,Jiraskova:2012:ufdl,nagy:2004:nfre:proc,nagy:2006:ufrlcfpa,Nagy:2019:ufdc,Nagy:2021:ufr}. In particular, in
their setting union-free languages are characterized by $1$CFPAs, that
is, nondeterministic finite automata having exactly one cycle-free
accepting path from every state. Furthermore, every regular language can
be represented as a finite union of union-free languages, giving rise to
the notion of union-complexity.

It should be noted, however, that the definition of union-free
languages used in these works differs from the unified framework adopted
in the present survey. In the classical definition, expressions are
built from single alphabet symbols by using concatenation and Kleene
star, but no union. In contrast, our framework allows an arbitrary
finite set of initial words $\Lambda$. Thus, finite sets of words may be
used as atomic building blocks. Consequently, languages such as
\[
\{b,c\} \{ac\}^*\{a,b\}
\]
or, more generally, expressions involving non-starred finite sets of words,
can be described in our formalism, although they do not belong to the
language family characterized by $1$CFPAs. In particular, every nonempty union-free language has a unique shortest word.

This distinction is not caused by starred finite sets. Indeed, if a
finite set of words occurs under a Kleene star, then it can be simulated
in the classical union-free setting. For example, a factor of the form
\[
\{w_0,w_1,\ldots,w_k\}^*
\]
can be represented union-freely as
\[
(\{w_0\}^*\{w_1\}^*\cdots\{w_k\}^*)^*.
\]

Hence, the essential difference between the two formalisms lies in the availability of finite sets of words that are not immediately placed under Kleene star. Our generalized definition is therefore slightly more liberal than the classical notion of union-free languages. Nevertheless, the absence of union still imposes substantial restrictions; for instance, the language
\[
\{\,a^m \mid m \geq 0\,\} \cup \{\,b^n \mid n \geq 0\,\}
\]
cannot be represented within our formalism.

Replacing Kleene star by complementation yields the class
$\mathcal{L}_{(\cdot,\compl{\phantom{m}})}$.

Although complementation allows to describe infinite languages, the resulting language family still forms a proper subclass of the star-free languages \cite{Brzozowski:1978:ddhsfli,cohen:1971:ddosfe}. A more detailed analysis of this subclass and its position within the dot-depth hierarchy would be of particular interest. In this context, it would also be worthwhile to investigate more systematically the role of union and to determine to what extent its absence is responsible for the structural and expressive restrictions of the resulting language family.

\begin{theorem}
The language family
$\mathcal{L}_{(\cdot,\compl{\phantom{m}})}$
is a proper subclass of the star-free languages.
\end{theorem}

A representative witness is the language
\[
(a\Sigma^*a)\cup(b\Sigma^*b),
\]
which is star free but cannot be represented using only concatenation
and complementation.

The class
$\mathcal{L}_{(*,\compl{\phantom{m}})}$
differs fundamentally from all previously considered classes. Since both
available operations are unary, expression trees degenerate into simple
paths. Every expression is therefore obtained by repeatedly alternating
Kleene star and complementation, starting from a finite language.
Naturally, this raises the question whether repeated alternation leads
to an infinite hierarchy, comparable to the dot-depth hierarchy of
star-free languages, or whether the process eventually stabilizes.

Surprisingly, the latter is the case.

\begin{theorem}
Every language in
$\mathcal{L}_{(*,\compl{\phantom{m}})}$
can be represented by an expression containing at most three
applications of Kleene star and three applications of complementation.
\end{theorem}

Equivalently, repeatedly alternating star and complementation produces
only finitely many essentially different forms before entering a
periodic cycle. Thus, unlike the dot-depth hierarchy, alternation
between these two unary operations does not increase expressive power
indefinitely.

The last nontrivial two-operation class is
$\mathcal{L}_{(\cap,*)}$.
Replacing complementation by intersection drastically changes the
structure of the generated languages. Since intersection only restricts
languages already obtained by Kleene star, every expression admits a
simple normal form.

\begin{theorem}
Every language in
$\mathcal{L}_{(\cap,*)}$
is either finite or can be represented as
\[
L_1^*\cap L_2^*\cap\cdots\cap L_k^*,
\]
where each $L_i$ is finite.
\end{theorem}

This characterization immediately implies several structural
restrictions. In particular, every infinite language in
$\mathcal{L}_{(\cap,*)}$
is closed under taking positive powers of its words. For unary
alphabets, the characterization becomes even simpler.

\begin{theorem}
Every unary language in
$\mathcal{L}_{(\cap,*)}$
is either finite or can be represented as $L^*$ for some finite language
$L$.
\end{theorem}

Hence, unlike the general case, intersections of starred finite unary
languages do not increase expressive power.

\subsection{Comparison of the Two-Operation Classes}

Having described the individual two-operation classes, we now compare
their expressive power. The resulting picture is highly non-linear. In
particular, over arbitrary alphabets none of the nontrivial
two-operation classes contains another one. Thus, apart from the classes
that collapse to the finite or finite/cofinite languages, the remaining
families are pairwise incomparable.

\begin{figure}[htbp]
\centering
\scalebox{0.9}{
\tikzset{
	box/.style={
		draw,
		rounded corners=3mm,
		minimum width=38mm,
		minimum height=10mm,
		align=center,
		fill=white,
		blur shadow
	}
}
	\begin{tikzpicture}[>=stealth]
		\node[box] (reg) at (0,1.2) {$\reg$};
		\node[box] (sc) at (0,-2.0)
		{$\mathscr{L}_{(*,\compl{\phantom{m}})}$};
		\node[box] (uf) at (-5,-3.6)
		{$\mathscr{L}_{(\cdot,*)}$};
		\node[box] (ccf) at (5,-3.6)
		{$\mathscr{L}_{(\cdot,\compl{\phantom{m}})}$};
		\node[box] (is) at (-3,-6.0)
		{$\mathscr{L}_{(\cap,*)}$};
		\node[box] (cf) at (3,-6.0)
		{$\mathscr{L}_{(\cup,*)}$};
		\node[box,minimum width=95mm] (fc)
		at (0,-9.2)
		{$
			\mathscr{L}_{(\cup,\compl{\phantom{m}})}
			=
			\mathscr{L}_{(\cap,\compl{\phantom{m}})}
			=\fin\cup\cofin
			$};
		\node[box,minimum width=95mm] (fin)
		at (0,-12.4)
		{$
			\mathscr{L}_{(\cup,\cdot)}
			=
			\mathscr{L}_{(\cup,\cap)}
			=
			\mathscr{L}_{(\cdot,\cap)}
			=
			\fin
			$};
%		
		% dashed separators
		\draw[dashed] (-6.5,-0.5) -- (6.5,-0.5);
		\draw[dashed] (-6.5,-7.6) -- (6.5,-7.6);
		\draw[dashed] (-6.5,-10.8) -- (6.5,-10.8);
%		
		% incomparability marker
		\node[circle,draw,dashed,minimum size=8mm] (inc) at (0,-4) {$\not\subseteq$};
		\draw[dashed] (sc) -- (inc);
		\draw[dashed] (uf) -- (inc);
		\draw[dashed] (ccf) -- (inc);
		\draw[dashed] (cf) -- (inc);
%		
		% proper inclusion: L_(cap,*) subseteq L_(dot,*)
		\draw[->]
		(is) to[bend right=10]
		node[midway,left]{\scriptsize conj.}
		(uf);
%		
		% unary inclusions
		\draw[->]
		(is) to[bend right=18]
		node[midway,above,xshift=-12mm,yshift=-10mm]{\scriptsize unary}
		(sc);
		\draw[->]
		(is) to[bend right=12]
		node[midway,below]{\scriptsize unary}
		(cf);
		\draw[->]
		(ccf) to[bend left=18]
		node[midway,above,xshift=5mm,yshift=-1mm]{\scriptsize unary}
		(sc);
		\draw[->]
		(ccf) to[bend left=18]
		node[midway,above,xshift=23mm,yshift=3mm]{\scriptsize unary}
		(uf);
		\draw[->]
		(ccf) to[bend left=18]
		node[midway,above,xshift=7mm,yshift=0mm]{\scriptsize unary}
		(cf);
	\end{tikzpicture}
}
\caption{Inclusion structure of the language families
$\mathcal{L}_{\Phi}$ induced by the different sets of two operations.
An arrow labeled \textsf{unary} denotes an inclusion that holds for
unary languages only. Over arbitrary alphabets, the nontrivial
two-operation classes are pairwise incomparable. So dashed lines mean incomparability.}
\label{fig:fam-inclusions-two}
\end{figure}
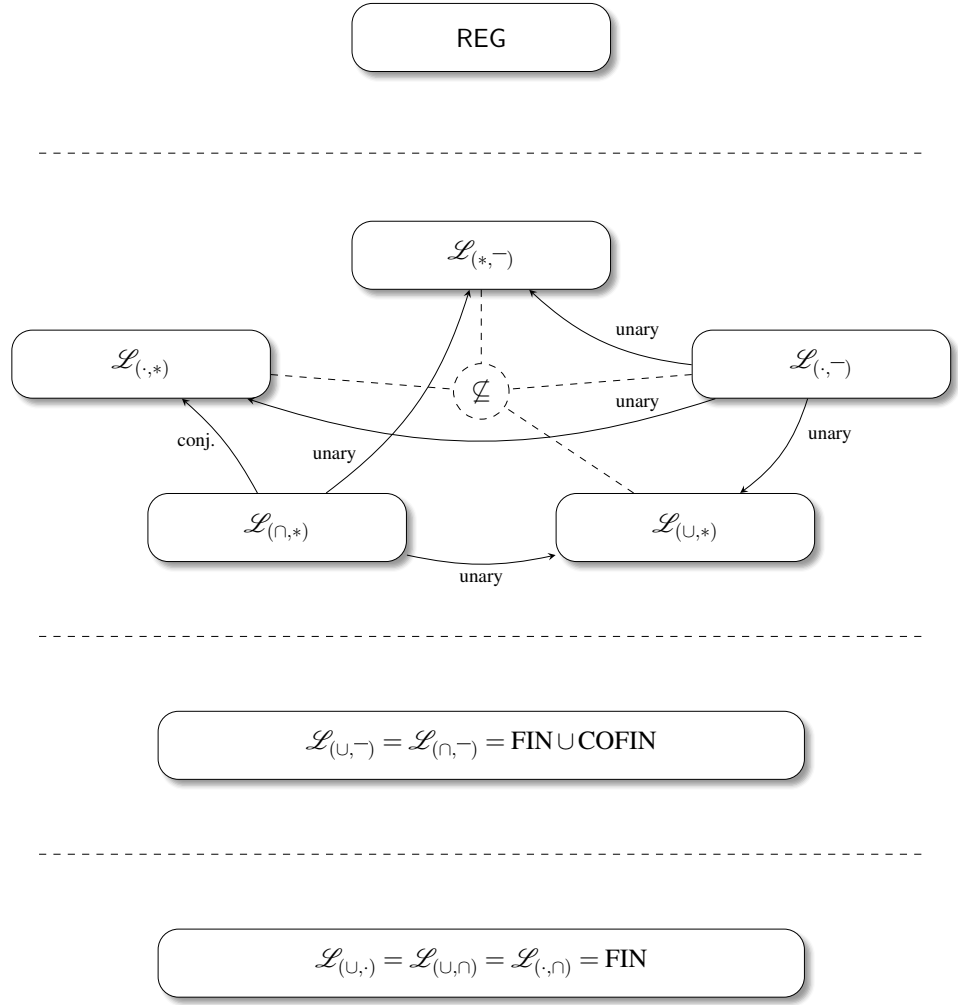

The first group of classes has already been identified above. The
families
\[
\mathcal{L}_{(\cup,\cdot)},\qquad
\mathcal{L}_{(\cup,\cap)},\qquad
\mathcal{L}_{(\cdot,\cap)}
\]
all coincide with the finite languages, while
\[
\mathcal{L}_{(\cup,\compl{\phantom{m}})}
=
\mathcal{L}_{(\cap,\compl{\phantom{m}})}
\]
coincides with the union of the families of finite and cofinite languages. These
classes therefore form the bottom part of the hierarchy. There is, however, already a language family outside this group that does not contain all cofinite languages, namely $\mathcal{L}_{(\cap,*)}$. Nevertheless, we place it higher in the hierarchy, since its expressive power extends beyond this narrow family of finite and cofinite languages.

The remaining five classes,
\[
\mathcal{L}_{(\cup,*)},\qquad
\mathcal{L}_{(\cdot,*)},\qquad
\mathcal{L}_{(\cdot,\compl{\phantom{m}})},\qquad
\mathcal{L}_{(\cap,*)},\qquad
\mathcal{L}_{(*,\compl{\phantom{m}})},
\]
behave quite differently. We conjecture that $\mathcal{L}_{(\cap,*)}$ is a proper subclass of $\mathcal{L}_{(\cdot,*)}$. Apart from this, the remaining classes are pairwise incomparable in general.

\begin{theorem}\label{theo:two-op-pairwise-incomp}
Over arbitrary alphabets, the language families
\[
\mathcal{L}_{(\cup,*)},\qquad
\mathcal{L}_{(\cdot,*)},\qquad
\mathcal{L}_{(\cdot,\compl{\phantom{m}})},\qquad
\mathcal{L}_{(*,\compl{\phantom{m}})}
\]
are pairwise incomparable.
\end{theorem}

The assertion of Theorem~\ref{theo:two-op-pairwise-incomp} comprises
ten pairwise comparisons. In order to keep the incomparability statements readable, we collect the
witness languages in Table~\ref{tab:witness-two}. Each row gives a
language that belongs to one of the five non-trivial two-operation
classes and, as indicated in the last column, does not belong to the
corresponding other classes. Thus, the pairwise incomparability results
can be read off from the table.

\begin{table}[!ht]
\centering
\renewcommand{\arraystretch}{1.25}
\begin{tabular}{|c|c|c|}
\hline
Class & Witness language & Separates from \\
\hline\hline

$\mathcal{L}_{(\cup,*)}$
&
$L_k=\{\,a^{km}\mid m\geq 0\,\}
 \cup
 \{\,b^{kn}\mid n\geq 0\,\}$
&
$\mathcal{L}_{(\cdot,*)}$,
$\mathcal{L}_{(\cdot,\compl{\phantom{m}})}$,
$\mathcal{L}_{(\cap,*)}$,
$\mathcal{L}_{(*,\compl{\phantom{m}})}$
\\
\hline
$\mathcal{L}_{(\cdot,*)}$
&
$\{\,ab^n\mid n\geq 1\,\}$
&
$\mathcal{L}_{(\cup,*)}$,
$\mathcal{L}_{(\cap,*)}$,
$\mathcal{L}_{(*,\compl{\phantom{m}})}$
$\mathcal{L}_{(\cdot,\compl{\phantom{m}})}$
$$
\\
\hline

\multirow{3}{*}{$\mathcal{L}_{(\cap,*)}$}
&
$L_1^*\cap L_2^*$
with
$L_1=\{baa,aa,ac\}$ and
$L_2=\{ba,aa,aac\}$
&
$\mathcal{L}_{(\cup,*)}$
\\

&
$\{\,(aa)^n\mid n\ge 0\,\}$
&
$\mathcal{L}_{(\cdot,\compl{\phantom{m}})}$
\\

&

$L_1^*\cap L_2^*$
with
$L_1=\{a,ab\}$ and
$L_2=\{aa,ab,ba,bb\}$

&
$\mathcal{L}_{(*,\compl{\phantom{m}})}$
\\
\hline

\multirow{2}{*}{$\mathcal{L}_{(\cdot,\compl{\phantom{m}})}$}
&
$\compl{a\Sigma^*a},\quad \Sigma=\{ a,b\}$
&
$\mathcal{L}_{(\cup,*)}$,
$\mathcal{L}_{(\cdot,*)}$,
$\mathcal{L}_{(*,\compl{\phantom{m}})}$
\\

&
$\{a, aaa\}\cup \{\,a^{n+42}\mid n\ge 0\,\}$
&
$\mathcal{L}_{(\cap,*)}$
\\
\hline

\multirow{2}{*}{$\mathcal{L}_{(*,\compl{\phantom{m}})}$}
&
$\compl{\{ \{aa,bb\}^n\mid n\ge 0\}}$
&
$\mathcal{L}_{(\cup,*)}$,
$\mathcal{L}_{(\cap,*)}$
\\

&
$\compl{\{\{a^6,a^9\}^n\mid n\ge 0\} }$
&
$\mathcal{L}_{(\cdot,*)}$,
$\mathcal{L}_{(\cdot,\compl{\phantom{m}})}$
\\

\hline
\end{tabular}
\caption{Witness languages for the pairwise incomparability of the five
non-trivial language families generated by two operations.}
\label{tab:witness-two}
\end{table}

The situation becomes more structured when we restrict the attention to
unary languages. While the four nontrivial language families are
pairwise incomparable over arbitrary alphabets, several proper
inclusions arise in the unary case.

A first observation is that, except for
$\mathcal{L}_{(\cap,*)}$, every two-operation language family
contains all finite and cofinite unary languages. Indeed,
$\mathcal{L}_{(\cup,\compl{\phantom{m}})}$
consists precisely of these languages, while
$\mathcal{L}_{(\cdot,\compl{\phantom{m}})}$
coincides with this class in the unary case. In detail,

\[
\mathcal{L}_{(\cup,\compl{\phantom{m}})}
=
\mathcal{L}_{(\cdot,\compl{\phantom{m}})}
\subset
\mathcal{L}_{(\cdot,*)},
\qquad
\mathcal{L}_{(\cdot,\compl{\phantom{m}})}
\subset
\mathcal{L}_{(\cup,*)},
\qquad
\mathcal{L}_{(\cdot,\compl{\phantom{m}})}
\subset
\mathcal{L}_{(*,\compl{\phantom{m}})}
\]
for unary languages.
The unary setting admits several inclusion results that do not hold over
arbitrary alphabets. In particular, the class
$\mathcal{L}_{(\cap,*)}$ is the weakest among the language families
containing Kleene star.

\begin{theorem}
For unary languages, the following strict inclusions hold:
\[
\mathcal{L}_{(\cap,*)}
\subset
\mathcal{L}_{(\cup,*)},
\qquad
\mathcal{L}_{(\cap,*)}
\subset
\mathcal{L}_{(\cdot,*)},
\qquad
\mathcal{L}_{(\cap,*)}
\subset
\mathcal{L}_{(*,\compl{\phantom{m}})}.
\]
All inclusions are proper.
\end{theorem}

Taken together, the unary case exhibits a considerably more structured
picture than the general case. Several language families that are
pairwise incomparable over arbitrary alphabets become related by proper
inclusions.

\section{Subregular Expressions with three Operations}

We now turn to the language classes generated by sets of three operations. Some of these classes can be identified immediately from the results established above. The class
$
\mathcal{L}_{(\cup,\cdot,\cap)}
$
contains only finite languages, since none of its operations can generate an infinite language from the finite literals. The class generated by the three Boolean operations,
$
\mathcal{L}_{(\cup,\cap,\compl{\phantom{m}})},
$
is only slightly more expressive. Starting from finite languages, complementation yields cofinite languages, and the family of finite and cofinite languages is closed under union, intersection, and complementation. Consequently,
$
\mathcal{L}_{(\cup,\cap,\compl{\phantom{m}})}
$
coincides with the class of finite and cofinite languages. Clearly
$
\mathcal{L}_{(\cup,\cdot,*)}
$
define the regular languages. The family $\mathcal{L}_{(\cup,\cdot,\compl{\phantom{m}})}$ is equal to $\mathcal{L}_{(\cap,\cdot,\compl{\phantom{m}})}$ as well as the family $\mathcal{L}_{(\cup,*,\compl{\phantom{m}})}$ is equal to $\mathcal{L}_{(\cap,\cdot,\compl{\phantom{m}})}$.

Apart from the equivalent classes this leaves five essentially different classes whose expressive power remains to be considered. The
class
$\mathcal{L}_{(\cup,\cdot,\compl{\phantom{m}})}$
of star-free languages has been studied extensively and constitutes one
of the best understood subclasses of the regular languages \cite{Brzozowski:1978:ddhsfli,cohen:1971:ddosfe,McNaughton:1971:cfa:book,Schuetzenberger:1965:ofmhots}. It admits
numerous characterizations in algebraic, logical, and automata-theoretic
terms. We begin with two of the remaining four classes, obtained by replacing union with either complementation or intersection while retaining concatenation and Kleene star. These are the classes
$\mathcal{L}_{(\cdot,*,\compl{\phantom{m}})}$ and
$\mathcal{L}_{(\cdot,*,\cap)}$. Finally, we consider the classes in which concatenation is omitted,
namely
$\mathcal{L}_{(\cup,*,\compl{\phantom{m}})}$ and
$\mathcal{L}_{(\cup,*,\cap)}$. 

We begin with the classes obtained by replacing the union operation
while retaining concatenation and Kleene star. The basic results can be found in \cite{kutrib:2018:ecse,werner:2013:eufsuua}. A natural question is whether these classes still capture all regular
languages. For both classes the answer is negative. The key observation
is that unary languages described by such expressions exhibit strong
structural restrictions. Exploiting this fact, it can
be shown that certain simple unary regular languages cannot be described
by $\mathcal{L}_{(\cdot,*,\cap)}$.

\begin{theorem}
For every integer $x\geq 2$, the unary language
\[
\{ a\}\cup \{\,a^n \mid n\equiv 0 \pmod x\,\}
\]
is not in $\mathcal{L}_{(\cdot,*,\cap)}$.
\end{theorem}

This result immediately implies that the family of unary languages in 
$\mathcal{L}_{(\cdot,*,\cap)}$ forms a proper subclass of the unary
regular languages. The underlying reason is the restricted way in which isolated words and infinite sets of words can be generated using only concatenation, Kleene star, and intersection. Kleene star always introduces an iterative structure: whenever a word $(a^k)$ is generated by a starred subexpression, all of its iterations $(a^{km})$, for $m\geq 0$, are generated as well. Thus, a word occurring through such an iteration cannot be isolated from its further repetitions.

Intersection cannot create a new exceptional word either. If a word belongs to an intersection $L_1\cap L_2$, then it must already belong to both $L_1$ and $L_2$. Hence, in order to retain the isolated word $a$, every intersected component contributing to the final expression must contain $a$. At the same time, the infinitely many words $a^{km}$, $m\geq 0$, have to arise from iterative subexpressions. The corresponding iteration mechanisms, however, cannot be restricted so as to preserve $a$ as a single exceptional word while generating exactly all multiples. Once $a$ participates in the relevant starred structure, its further iterations are generated as well, whereas intersection can only remove words by requiring membership in all component languages and cannot introduce $a$ independently of these iterative patterns. This observation captures the main intuition behind this language class.

The situation differs markedly for union-free
languages. A result in \cite{werner:2013:eufsuua} shows that every unary regular language is
in $\mathcal{L}_{(\cdot,*,\compl{\phantom{m}})}$. Consequently, one obtains the following strict inclusion.

\begin{corollary}
The family of unary languages in $\mathcal{L}_{(\cdot,*,\cap)}$ is strictly
included in the family of unary languages in $\mathcal{L}_{(\cdot,*,\compl{\phantom{m}})}$.
\end{corollary}

Whether the same inclusion remains proper for arbitrary alphabets is
currently an open problem. Nevertheless, $\mathcal{L}_{(\cdot,*,\compl{\phantom{m}})}$ does not
coincide with the regular languages. Although every regular language can
be represented as a finite union of languages from $\mathcal{L}_{(\cdot,*,\compl{\phantom{m}})}$.

\begin{theorem}
The language
\[
L=\{\lambda \}
\cup \{\,awa\mid w\in\{a,b\}^*\,\}
\cup \{\,bwb\mid w\in\{a,b\}^*\,\}
\]
is not in $\mathcal{L}_{(\cdot,*,\compl{\phantom{m}})}$.
\end{theorem}

The basic intuition is that generating arbitrarily long blocks consisting only of $a$'s or only of $b$'s requires the use of Kleene star. However, Kleene star also introduces unrestricted iteration: once a subexpression generates a valid nonempty word, further copies of words generated by the same starred subexpression may be concatenated. This makes it difficult to preserve the requirement that the first and last symbols of a word coincide. For example, a valid word beginning and ending in $a$ may be followed by a valid word beginning and ending in $b$, producing a word whose first and last symbols differ and which therefore does not belong to $L$. Complementation does not resolve this problem. Although it can exclude certain sets of words, it does not provide a mechanism for coordinating the first and last symbols across arbitrarily long concatenations generated by starred subexpressions, it can only build the complement of stared and non-starred sets.

As a consequence, the language family $\mathcal{L}_{(\cdot,*,\compl{\phantom{m}})}$ forms a proper subclass of the regular languages.

The next classes to be considered are
$\mathcal{L}_{(\cup,*,\cap)}$ and
$\mathcal{L}_{(\cup,*,\compl{\phantom{m}})}$,
which arise by removing concatenation. In contrast to the previous
setting, the absence of concatenation makes the finite basis
$\Lambda$ indispensable. Without concatenation, words can no longer be
constructed from individual alphabet symbols, and therefore the
expressions operate directly on the finite collection of initial words.
This explains the generalized framework
$\regex(\Sigma,\Lambda,\Phi)$ introduced earlier.

These concatenation-free language families have been investigated in \cite{kutrib:2015:eccf:proc,kutrib:2017:cfl,kutrib:2018:ecse}. Their expressive power turns
out to be considerably more limited than that of classes containing
concatenation.

The absence of concatenation imposes strong structural restrictions on
the unary languages that can be described. A complete characterization
can be obtained in terms of languages that are either finite or arise
from cofinite unary languages by stretching all word lengths by a fixed
factor. 
More precisely, let $m\geq 1$ be an integer.
A \emph{unary} language $L\subseteq \{a\}^*$ is stretched by $m$ to 
a language $L_{(m)}$ in the following sense:
$L_{(m)}=\{\,a^{m\cdot n}\mid a^{n}\in L\,\}$. That is, a unary language
is stretched by multiplying each word length by $m$. We say that
language $L_{(m)}$ \emph{is obtained from $L$ stretched by $m$}.
The family of all unary languages that are either finite or
are obtained from a cofinite language stretched by some $m\geq 1$ and including the 
empty word is denoted by $\ufam$.  We are particularly interested in the
union closure~$\Gamma_\cup(\ufam)$ of $\ufam$.
Each language $L \in \Gamma_\cup(\ufam)$ has a representation
$$
\bigcup_{1\leq i\leq k} L_{i},
\text{ where } k\geq 0 \text{ and } L_{i} \in \ufam.
$$

The key observation is that $\Gamma_\cup(\mathcal U)$ is closed under
all operations available in $\mathcal{L}_{(\cup,*,\cap)}$. Closure under union is immediate by definition. Moreover,
the family is closed under intersection, since the intersection of two
stretched cofinite languages is again of the same form, up to finitely
many exceptions. Finally, applying Kleene star to a language from
$\Gamma_\cup(\mathcal U)$ again yields a language of the same type.

This leads to a complete characterization of the unary
$\mathcal{L}_{(\cup,*,\cap)}$.

\begin{theorem}
A unary language is in $\mathcal{L}_{(\cup,*,\cap)}$ if and only if it
belongs to $\Gamma_\cup(\mathcal U)$.
\end{theorem}

The characterization immediately reveals limitations of the expressive
power of $\mathcal{L}_{(\cup,*,\cap)}$. In particular,
certain unary periodic languages cannot be represented.

\begin{theorem}
For integers $1\leq x<y$, the language
\[
L=\{\, a^n \mid n\equiv x \pmod y\,\}\cup\{\lambda \}
\]
is not in $\mathcal{L}_{(\cup,*,\cap)}$.
\end{theorem}

Since complementation is available in
$\mathcal{L}_{(\cup,*,\compl{\phantom{m}})}$,
intersection can be simulated by De Morgan's laws.
Consequently, every language in $\mathcal{L}_{(\cup,*,\cap)}$ is in
$\mathcal{L}_{(\cup,*,\compl{\phantom{m}})}$. The converse, however, does not hold.

\begin{theorem}
The class of unary languages in $\mathcal{L}_{(\cup,*,\cap)}$ is strictly
included in the class of unary languages in $\mathcal{L}_{(\cup,*,\compl{\phantom{m}})}$.
\end{theorem}

A useful characterization of the unary case of $\mathcal{L}_{(\cup,*,\compl{\phantom{m}})}$ is obtained through the Boolean closure of $\ufam$. Let $\ukf$ denote the least family of languages containing $\ufam$ and closed under union and complementation, and hence also under intersection. Equivalently, every language $L\in\ukf$ can be represented in the form
$$
\bigcup_{1\leq i\leq k} \bigcap_{1\leq j\leq l_i} L_{i,j},
\text{ where } k,l_1,l_2,\dots, l_k\geq 0 \text{ and }
L_{i,j} \in \ufam \text{ or } \compl{L}_{i,j} \in \ufam.
$$

This Boolean closure provides an exact characterization of unary concatenation-free languages.

\begin{theorem}\label{theo}
A unary language is in $\mathcal{L}_{(\cup,*,\compl{\phantom{m}})}$ if and only if it belongs to $\ukf$, that is, to the Boolean closure of $\ufam$.
\end{theorem}

Thus, in the unary setting, allowing Boolean combinations of languages from $\ufam$ captures precisely the expressive power of concatenation-free expressions. This characterization also provides a convenient basis for showing that languages cannot be expressed.

\subsection{Comparison of the Three-Operation Classes}
Having considered the individual language families generated by sets of three operations, we now compare their expressive power and determine the inclusion and incomparability relations between them.

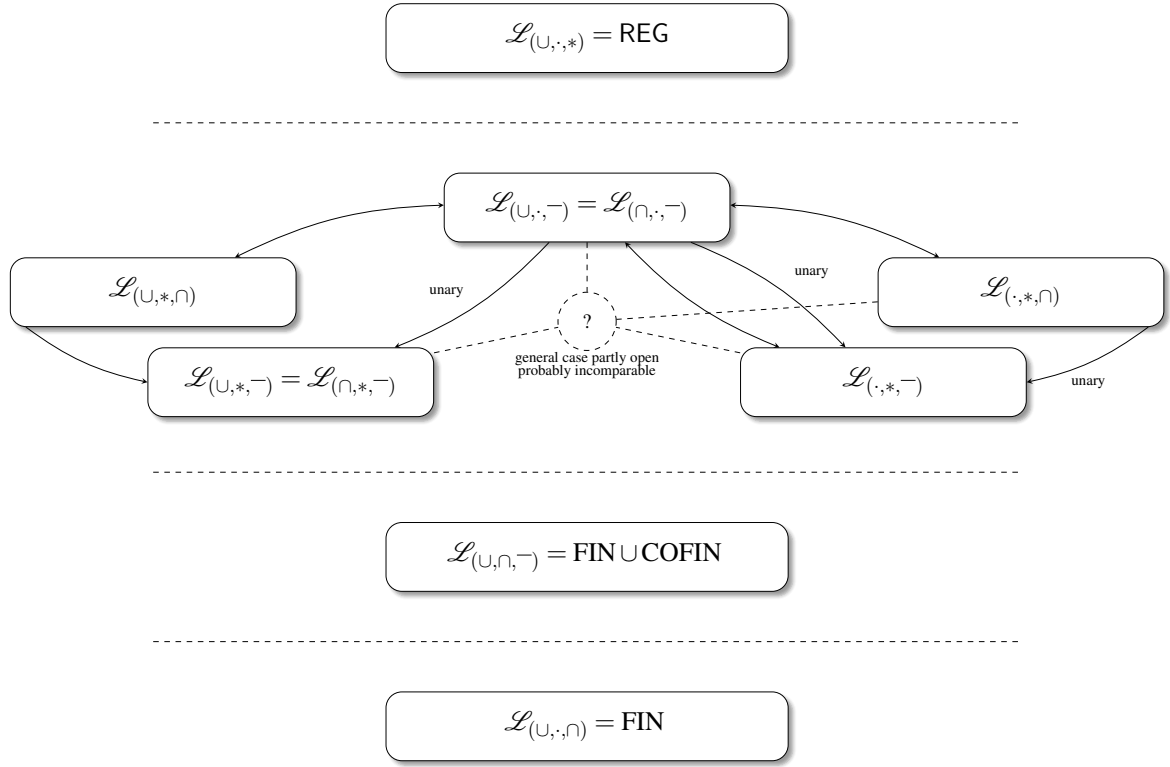
\begin{figure}
\centering
\scalebox{0.7}{
% larger font for language classes
\newcommand{\langclass}[1]{{\Large $\displaystyle #1$}}
\tikzset{
	box/.style={
		draw,
		rounded corners=3mm,
		minimum width=44mm,
		minimum height=10mm,
		align=center,
		fill=white,
		blur shadow
	},
	widebox/.style={
		draw,
		rounded corners=3mm,
		minimum width=76mm,
		minimum height=13mm,
		align=center,
		fill=white,
		blur shadow
	},
	region/.style={
		draw,
		rounded corners=18mm,
		inner sep=8mm,
		fill=gray!8
	},
	midcircle/.style={
		circle,
		draw,
		dashed,
		minimum size=11mm,
		align=center,
		fill=white
	}
}
\begin{tikzpicture}[>=stealth]
	% REG
	\node[widebox] (reg) at (0,3.8)
	{\langclass{
			\mathscr{L}_{(\cup,\cdot,*)}
			=
			\reg
	}};
	% upper circular arrangement
	\node[widebox,minimum width=54mm,minimum height=13mm] (cf) at (-5.6,-2.7)
	{\langclass{
			\mathscr{L}_{(\cup,*,\compl{\phantom{m}})}
			=
			\mathscr{L}_{(\cap,*,\compl{\phantom{m}})}
	}};
	\node[widebox,minimum width=54mm,minimum height=13mm] (sf) at (0,0.6)
	{\langclass{
			\mathscr{L}_{(\cup,\cdot,\compl{\phantom{m}})}
			=
			\mathscr{L}_{(\cap,\cdot,\compl{\phantom{m}})}
	}};
	\node[box,minimum width=54mm,minimum height=13mm] (uf) at (5.6,-2.7)
	{\langclass{
			\mathscr{L}_{(\cdot,*,\compl{\phantom{m}})}
	}};
	\node[box,minimum width=54mm,minimum height=13mm] (ufi) at (8.2,-1.0)
	{\langclass{
			\mathscr{L}_{(\cdot,*,\cap)}
	}};
	\node[box,minimum width=54mm,minimum height=13mm] (cfi) at (-8.2,-1.0)
	{\langclass{
			\mathscr{L}_{(\cup,*,\cap)}
	}};	
	\node[midcircle] (inc) at (0,-1.55)
	{$?$};	
	% known proper inclusion
	\draw[->]
	([xshift=3mm]cfi.south west) to[bend right=12]
	%node[midway,left,xshift=-2mm,yshift=-1mm]{\scriptsize proper}
	(cf.west);	
	\draw[<->]
	(uf) to[bend left=12] (sf);	
	\draw[<->]
	(ufi) to[bend right=12] (sf);	
	\draw[<->]
	(cfi) to[bend left=12] (sf);	
	\draw[->]
	([xshift=-3mm]ufi.south east) to[bend left=12]
	node[midway,right,xshift=-5mm,yshift=-4mm]{\scriptsize unary}
	(uf.east);
%	
%	
	% unary known inclusions
	\draw[->]
	(sf) to[bend left=12]
	node[midway,above,xshift=-6mm]{\scriptsize unary}
	(cf);
	\draw[->]
	(sf) to[bend left=12]
	node[midway,above,xshift=7mm]{\scriptsize unary}
	(uf);
	\node at (0,-2.25) {\scriptsize general case partly open};
	\node at (0,-2.5) {\scriptsize probably incomparable};
%	
	% lower regions
	\node[widebox] (cofin) at (0,-6)
	{
		\langclass{
			\mathscr{L}_{(\cup,\cap,\compl{\phantom{m}})} = \fin\cup\cofin
		}
		%\\[1mm]
		%finite or cofinite
	};
	\node[widebox] (fin) at (0,-9.2)
	{
		\langclass{
			\mathscr{L}_{(\cup,\cdot,\cap)}
			=
			\fin
		}
	};
%	
	% unclear / presumed incomparability in the general case
	\draw[dashed] (sf) -- (inc);
	\draw[dashed] (cf) -- (inc);
	\draw[dashed] (uf) -- (inc);
	\draw[dashed] (ufi) -- (inc);
%	
	% lower separators
	\draw[dashed] (-8.2,2.2) -- (8.2,2.2);
	\draw[dashed] (-8.2,-4.4) -- (8.2,-4.4);
	\draw[dashed] (-8.2,-7.6) -- (8.2,-7.6);
\end{tikzpicture}
}
\caption{Inclusion structure of the language families
$\mathcal{L}_{\Phi}$ induced by the different sets of three
operations. Arrows with two heads
indicate incomparability, whereas arrows with one head indicate strict
inclusion. An arrow labeled \textsf{unary} denotes a relation that has
only been established for unary languages; the corresponding relation
for arbitrary alphabets may remain open. The dashed arrow indicates an
relation which is currently unknown.}
\label{fig:fam-inclusions}
\end{figure}

The relationships between the main language classes generated by three
operations are summarized in Table~\ref{tab:witness-three} and the
relation between the different language families in the
Figure~\ref{fig:fam-inclusions}. The table collects the witness languages used
to establish separations, together with known inclusion relations and cases
that remain open. It therefore provides an overview of the expressive
landscape before the individual relations are discussed in more detail.

\begin{table}[!ht]
\centering
\renewcommand{\arraystretch}{1.25}
\begin{tabular}{|c|c|c|}
\hline
Class & Witness language / known relation & Separates from / status \\
\hline\hline

$\mathcal{L}_{(\cup,\cdot,\compl{\phantom{m}})}$
&
$\{\lambda\}
 \cup
 \{\,awa\mid w\in\{a,b\}^*\,\}
 \cup
 \{\,bwb\mid w\in\{a,b\}^*\,\}$
&
$\mathcal{L}_{(\cdot,*,\cap)}$, $\mathcal{L}_{(\cdot,*,\compl{\phantom{m}})}$
\\
\cline{2-3}

&
$\{a\}\cdot\compl{\emptyset}$
&
$\mathcal{L}_{(\cup,*,\cap)}$
\\
\cline{2-3}

&

\text{general case open}
&
$\mathcal{L}_{(\cup,*,\compl{\phantom{m}})}$
\\
\cline{2-3}
&
\text{proper inclusion in the unary case}
&
$\mathcal{L}_{(\cup,*,\compl{\phantom{m}})}$
\\
\hline

$\mathcal{L}_{(\cup,*,\cap)}$
&
$\{\,(aa)^n\mid n\ge 0\,\}$
&
$\mathcal{L}_{(\cdot,*,\compl{\phantom{m}})}$
\\
\cline{2-3}

&
$\{ a\}\cup \{\, a^{5n}\mid n\ge 0\,\}$
&
$\mathcal{L}_{(\cdot,*,\cap)}$, $\mathcal{L}_{(\cdot,*,\compl{\phantom{m}})}$
\\
\hline

$\mathcal{L}_{(\cup,*,\compl{\phantom{m}})}$
&
$\{\,(aa)^n\mid n\ge 0\,\}$
&
$\mathcal{L}_{(\cdot,*,\compl{\phantom{m}})}$
\\
\cline{2-3}

&
$\{ a\}\cup \{\, a^{5n}\mid n\ge 0\,\}$
&
$\mathcal{L}_{(\cdot,*,\cap)}$, $\mathcal{L}_{(\cdot,*,\compl{\phantom{m}})}$
\\
\hline

$\mathcal{L}_{(\cdot,*,\cap )}$
&
$\{\,a^mb^n\mid m,n\ge 0\,\}$
&
$\mathcal{L}_{(\cup,\cdot,\compl{\phantom{m}})}$
\\
\cline{2-3}

&
$\{\,ab^{n}\mid n\geq 0\,\}$
&
$\mathcal{L}_{(\cup,*,\cap)}$,$\mathcal{L}_{(\cup,*,\compl{\phantom{m}})}$
\\
\hline

$\mathcal{L}_{(\cdot,*,\compl{\phantom{m}}))}$
&
$\{\,a^mb^n\mid m,n\ge 0\,\}$
&
$\mathcal{L}_{(\cup,\cdot,\compl{\phantom{m}})}$
\\
\cline{2-3}

&
$\{\,ab^{n}\mid n\geq 0\,\}$
&
$\mathcal{L}_{(\cup,*,\cap)}$,$\mathcal{L}_{(\cup,*,\compl{\phantom{m}})}$
\\
\hline

\end{tabular}
\caption{Witness languages and open relations for the main language
families generated by three operations. Equalities induced by De Morgan's
laws are included in the first column. Several inclusions are known to be
proper for unary languages, while the corresponding relations over
arbitrary alphabets remain open.}
\label{tab:witness-three}
\end{table}

\section{Conclusion and Open Problems}

We have surveyed language families induced by subregular expressions
obtained from restricted sets of regularity-preserving operations. The
resulting picture shows that even small changes in the available
operators may lead to substantial differences in expressive power. Some
operator sets collapse to well-known classes such as the finite
languages, the union of finite and cofinite languages, or the star-free languages. Other combinations give rise to
proper subclasses of the regular languages with rather specific
structural restrictions.

A central observation is that the hierarchy of these language families
is far from linear. While several inclusions hold in the unary case, the
corresponding classes often become incomparable over arbitrary
alphabets. Thus, the expressive power of a restricted expression system
depends not only on the number of available operations, but also on the
interaction between the particular operations and the structure of the
underlying alphabet. In this sense, subregular expressions provide a
fine-grained view on the internal landscape of the regular languages.

Several questions remain open for the classes generated by
three operations. For unary languages, it is known that
$
\mathcal{L}_{(\cdot,*,\cap)}
\subset
\mathcal{L}_{(\cdot,*,\compl{\phantom{m}})}
$
is a strict inclusion. Hence, over a unary alphabet, replacing
intersection by complementation in the presence of concatenation and
Kleene star strictly increases expressive power. However, the
corresponding situation over arbitrary alphabets is not yet fully
understood. It remains open whether
$
\mathcal{L}_{(\cdot,*,\cap)}
\subset
\mathcal{L}_{(\cdot,*,\compl{\phantom{m}})}
$
holds in general or whether the two classes are incomparable.

A similar phenomenon occurs for the classes
$
\mathcal{L}_{(\cup,\cdot,\compl{\phantom{m}})}
\text{ and }
\mathcal{L}_{(\cup,*,\compl{\phantom{m}})}.
$
In the unary case, star-free languages coincide with the union of finite and
cofinite unary languages, and they form a proper subclass of
$\mathcal{L}_{(\cup,*,\compl{\phantom{m}})}$. Thus, in the unary
setting, replacing concatenation by Kleene star in the presence of union
and complementation increases expressive power. For arbitrary alphabets,
however, the exact relationship between these two classes remains open.
In particular, it is not known whether the unary inclusion extends to
larger alphabets or whether the classes become incomparable.

These open problems show that the unary case, while often more
accessible, does not automatically determine the general situation.
Unary languages are governed by arithmetic properties of word lengths,
whereas languages over larger alphabets may additionally depend on the
order and interaction of different letters. Understanding which unary
inclusions lift to arbitrary alphabets, and which fail because of
non-unary phenomena, is therefore a central task for a complete
classification of the three-operation families.

A second direction for further research concerns
\emph{decidability questions}. Given a regular language, for instance by
a deterministic finite automaton, one may ask whether it can be described
by an expression over a prescribed set of operations. This
problem is well understood for some classical classes, such as
star-free languages, but remains open for
several of the restricted expression systems discussed here. More
generally, one may ask whether there exists an effective procedure that,
given a regular language~$L$ and an operator set $\Phi$, decides
whether $L\in\mathcal{L}_{\Phi}$. If the answer is positive, a second
natural question is whether such an expression can actually be
constructed.

Closely related to this is the search for further
\emph{structural characterizations}. For star-free languages,
problems can be studied using algebraic, logical, and
automata-theoretic tools. Similar characterizations would be highly
desirable for the remaining subregular families. Possible candidates
include conditions on the syntactic monoid, restrictions on minimal
deterministic automata, special forms of transition graphs, closure
properties, or normal forms for expressions. Such
criteria would not only clarify the expressive power of the individual
families, but also provide practical methods for deciding whether a
given language belongs to one of them.

Another important direction concerns \emph{descriptional complexity}.
Even if a language belongs to a restricted family, it may require a much
larger expression than in the unrestricted regular setting. Conversely,
some restricted systems may describe certain languages more succinctly
than classical regular expressions by using operations such as
intersection or complementation. This raises a number of natural
questions: How large must an expression over a given operator set be in
comparison with an equivalent classical regular expression? How does the
size of such an expression relate to the number of states of a minimal
DFA? Are there exponential or even nonelementary gaps between different
operator sets? And how does the complexity change if the number of
occurrences of a particular operation is bounded?

These questions suggest several refined complexity measures. Besides
the overall size of an expression, one may consider the number of
occurrences of union, concatenation, Kleene star, intersection, or
complementation, the nesting depth of unary operations, the star height,
or the size of the finite basis~$\Lambda$. Such measures could lead to
hierarchies inside the language families themselves, analogous to
classical hierarchies such as star height or dot depth. Some of the
results surveyed here already indicate that bounded resources.

Overall, subregular expressions offer a systematic way to study how
individual operations contribute to the expressive power of regular
expressions. A complete theory of these
restricted expression systems would therefore not only refine our
understanding of regular languages, but also provide new tools for
comparing language descriptions under limited resources.

\end{document}